\documentclass[%
 reprint,
nofootinbib,
nobibnotes,
 amsmath,amssymb,
 aps,
prb,
floatfix,
]{revtex4-2}

\usepackage{graphicx}
\usepackage{dcolumn}
\usepackage{bm}
\usepackage{soul}

\DeclareGraphicsExtensions{.pdf,.jpg, .png}
\usepackage{epstopdf}
\usepackage{enumitem}

\begin{document}
\preprint{}

\title{Ghost sensing: the rise and role of exceptional points in planar geometry
}


\author{Emroz Khan and Evgenii E. Narimanov}
\address{Elmore Family School of Electrical and Computer Engineering and Birck Nanotechnology Center, Purdue University, West Lafayette, Indiana 47906, United States}

\date{\today}
\begin{abstract}
We show the recently discovered ghost waves - a special class of non-uniform electromagnetic waves in biaxial anisotropic media - can be used for optical sensing based on exceptional points. In addition to showing high sensitivity and precision, the proposed sensor employs simple planar geometry and is robust against noise.
\end{abstract}



\maketitle

Despite decades of research in the field of optical sensing, which spans many areas of medical, environmental and research applications, detection of small analytes at ultra-low concentration remains challenging \cite{mejia2018plasmonic}. The widely used sensing methods based on surface plasmon resonance suffer from low quality factor of the detected signal owing to metallic loss, while sensors based on dielectric waveguides, on the other hand, require sophisticated interferometric techniques with non-trivial device geometries \cite{duval2015optical}.  \\

A recent line of research promises enhanced sensitivity though operation based on exceptional points \cite{makris2008beam, klaiman2008visualization, guo2009observation,  ruter2010observation, feng2017non, el2019dawn, miri2019exceptional}, which are spectral singularities present in open systems described by non-Hermitian Hamiltonians \cite{bender1998real,bender2007making}. Near an exceptional point, the resonant frequencies of a system show strong dependence on the external perturbations, which has led to the concept of exceptional point-based enhanced sensing in photonics \cite{wiersig2014enhancing,liu2016metrology,ren2017ultrasensitive,sunada2017large} and electronics \cite{chen2018generalized, dong2019sensitive}. Although increased sensitivity has been demonstrated in several systems \cite{chen2017exceptional,hodaei2017enhanced,lai2019observation,hokmabadi2019non}, the resulting devices usually require non-trivial microscale geometry, high-Q resonators and precise control of optical loss and gain -- with a resulting complexity that so far prevented the use of these novel devices in practical biosensing. \\

In this work, we solve this complexity problem by introducing an exceptional point-based sensor that employs ghost waves, a new kind of surface wave that has recently been introduced theoretically in \cite{narimanov2019ghost} and \cite{narimanov2018dyakonov}, and studied experimentally in \cite{ma2021ghost} at the interface between a biaxial anisotropic and an isotropic dielectric medium. Since these waves can be excited at the planar surface of bulk crystals in as simple as in a Kretchmann configuration, the need for complicated fabrication or doping is lifted. The resulting ``ghost sensor" which derives its sensing enhancement through exceptional point-based operation and its device simplicity through manipulation of ghost waves in bulk optics, has the potential to revolutionize biosensing technology.\\

\begin{figure}[tb]
\centering
   \includegraphics[width=2.5 in]{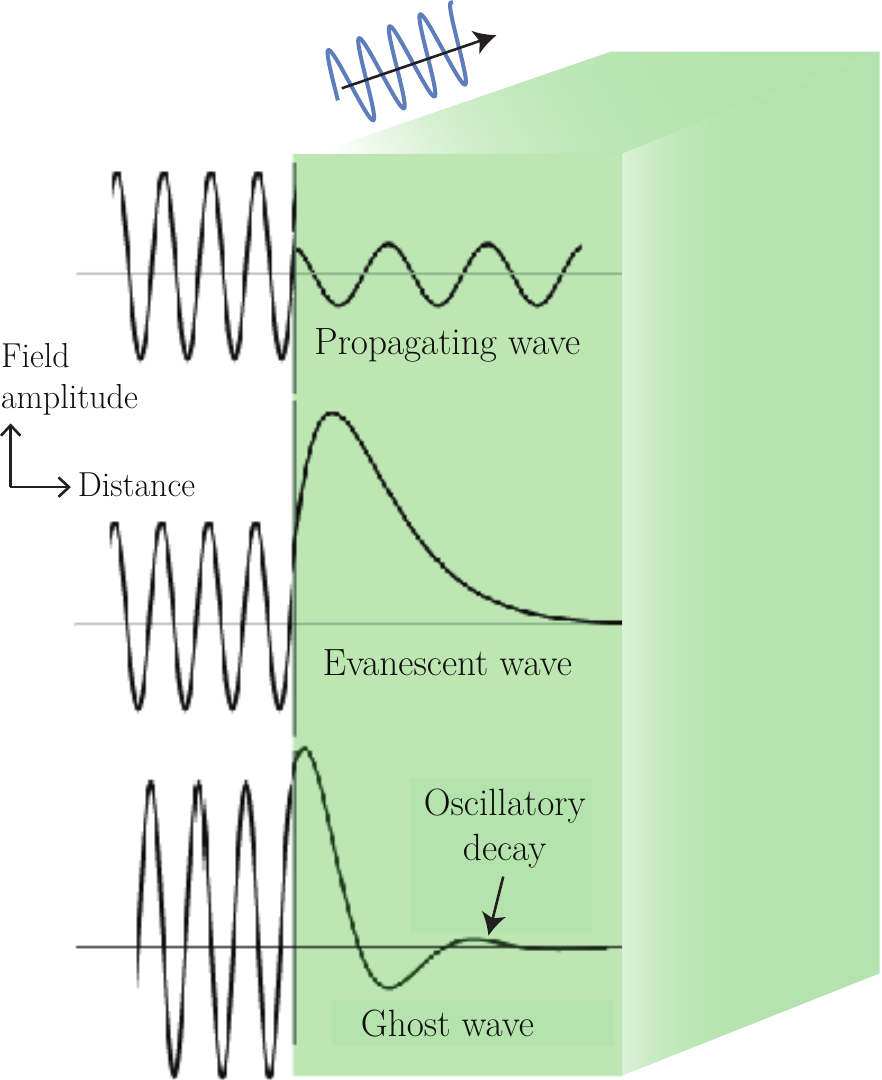}
\caption{\footnotesize A schematic of the interface between a biaxial anisotropic (green) and isotropic dielectric (white) shows that in addition to propagating and evanescent waves, a biaxial material can support ghost wave which decays with oscillations. This is in contrast with conventional Dyakonov surface wave which decays evanescently on both sides \cite{d1988new,takayama2009observation}.}
\label{Fig:waves}
\end{figure}

Ghost waves \cite{narimanov2019ghost} are a special class of non-uniform electromagnetic waves in biaxial anisotropic media that combine the properties of propagating and evanescent fields. As shown in Fig.\ref{Fig:waves}, within the biaxial medium these waves decay exponentially  with oscillations. This hybrid behavior of propagation and decay does not result from energy loss due to absorption or scattering, and ghost waves can be excited in a completely lossless system, such as the interface between a biaxial and isotropic dielectric. The designation ``ghost"  \cite{narimanov2019ghost} refers to the physical origin of these waves as they  are created in tangent bifurcations that annihilate pairs of positive and negative index modes and represent the optical analogue of the well known  ``ghost orbits" in the semiclassical quantization of non-integrable dynamical systems \cite{kus1993prebifurcation}.\\
 
The oscillatory character of ghost surface waves leads to new type of mode interaction \cite{debnath2021ghost} which can give rise to frequency degeneracies in the form of exceptional points. When two ghost waves interact through a spatial overlap of their fields, the oscillations in the exponential tail qualitatively changes the nature of coupling between the two modes. If the extent of field overlap is gradually varied by, for example, controlling the physical distance between the two waves, the resulting coupling also shows oscillations instead of a purely monotonic behavior. This leads to oscillations in the frequency splitting of the system's eigenmodes as well. In particular, for a certain value of the control parameter, the coupling can be made effectively zero which results in the two eigen-frequencies being degenerate. If the system is non-conservative, then the degeneracy gives rise to exceptional points where both the eigen-frequencies as well as the eigenmodes coalesce. \\

\begin{figure}[tb]
\centering
   \includegraphics[width=3.1 in]{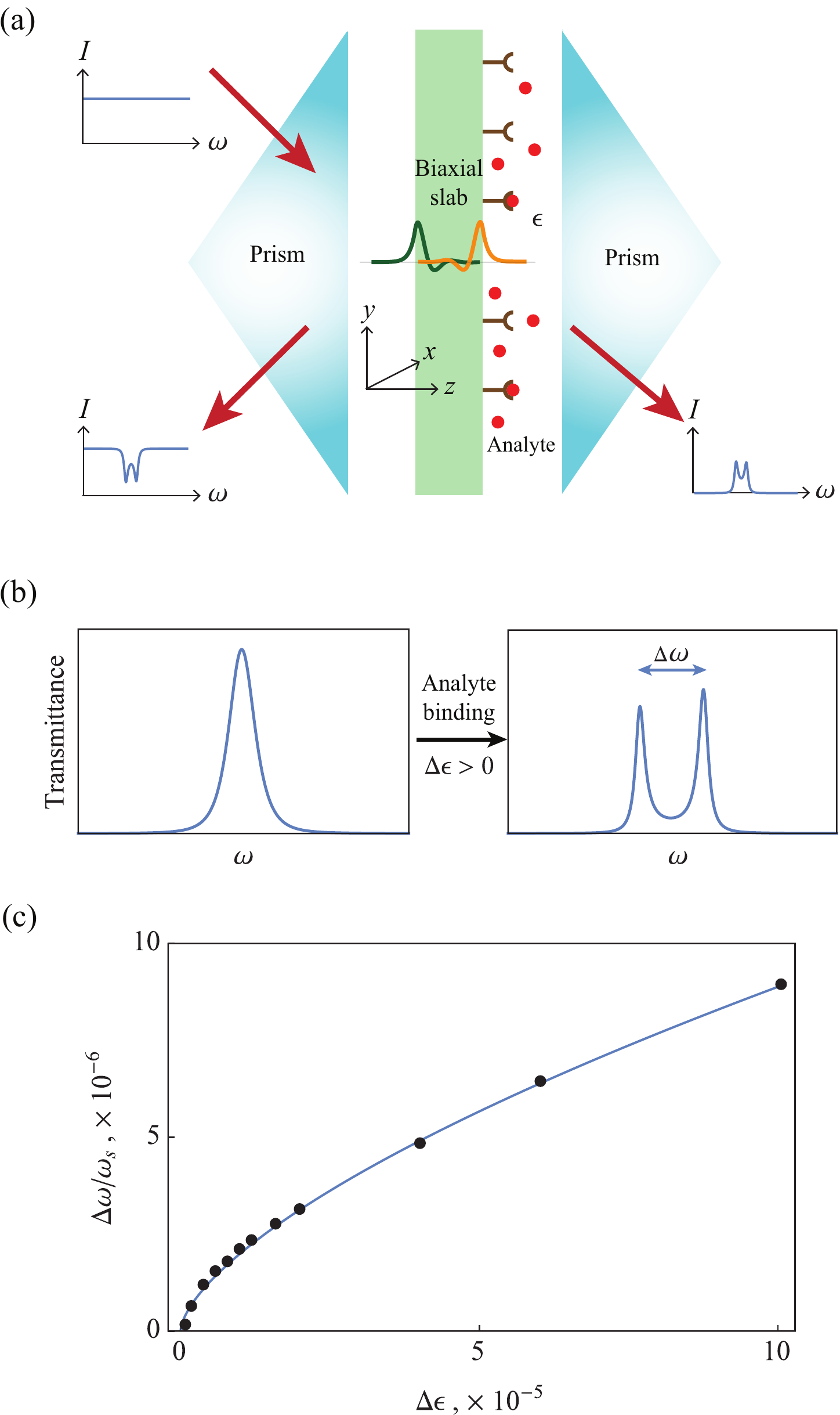}
\caption{\footnotesize Ghost sensing employs planar geometry and shows high sensitivity due to operation at an exceptional point. (a) The sensor includes a biaxial slab (e.g., NaNO$_2$ with permittivity tensor $\overline{\overline{\epsilon}}_{\rm slab} = {\rm diag} (2.726,1.806,1.991)$) surrounded by isotropic media (of permittivity $\epsilon \simeq 2.01$) and prism couplers on either sides. As the sensor is initially configured at an exceptional point, the frequencies to excite ghost surface waves on the two interfaces of the biaxial medium are degenerate. (b) Depending on permittivity changes on one of the isotropic sides due to binding of analytes, the frequencies split which can be measured from the transmission spectrum. (c) The amount of resonance splitting $\Delta \omega$ (sensor output) as a function of permittivity changes $\Delta \epsilon$ (sensor input) shows strong nonlinear behavior for small perturbation which is characteristic of exceptional point-based sensing. Here the fitted line corresponds to $\Delta\omega \propto (\Delta\epsilon)^{0.65}$. The thickness used for the biaxial slab is $12 \lambda_s$ and those for the two isotropic layers are $10 \lambda_s$ (left) and $4.05 \lambda_s$ (right) with $\lambda_s = 2 \pi c / \omega_s$ corresponding to ghost resonance wavelength for the single interface between biaxial (NaNO$_2$) and isotropic ($\epsilon = 2.01$) media. The in-plane momenta are $(q_x,q_y) \simeq (1.182,0.793) \cdot \omega_s/c$.}
\label{Fig:sensor}
\end{figure}

A sensor which can operate at such an exceptional point is schematically shown in Fig. \ref{Fig:sensor}(a) which includes a planar slab of biaxial material surrounded on both sides by a layer of isotropic medium followed by a prisms-coupler. Light incident through one of the prisms with a fixed tangential momentum would excite ghost surface waves at each of the two interfaces of the biaxial material. Owing to the finite thickness of the biaxial slab, the fields will spatially overlap, and in general, the corresponding excitation frequencies will be different. However, the permittivities and thicknesses of the two isotropic layers, which can be regarded as system parameters having a direct control over the amount of mode interaction as well as coupling to the environment, can be carefully chosen, as we explain shortly, such that the system is initially set up at an exceptional point -- leading to degeneracy in the excitation frequencies of ghost waves. \\

During a sensing operation a slight perturbation in the refractive index of one of the isotropic sides caused by binding of analyte would break the frequency degeneracy and the two resonances will diverge away from each other as can be seen in the transmission  (see Fig. \ref{Fig:sensor}(b)) or reflection spectrum. By measuring the distance between the two resonance frequencies, one can detect the presence of analytes and quantitatively determine its density. As can be seen from Fig. \ref{Fig:sensor}(c), which shows the amount of frequency splitting as a function of change in permittivity, the sensor exhibits strong nonlinear response for small perturbation which is characteristic of exceptional point-based sensing \cite{chen2017exceptional}.\\

\begin{figure}[tb]
\centering
   \includegraphics[width=3.3 in]{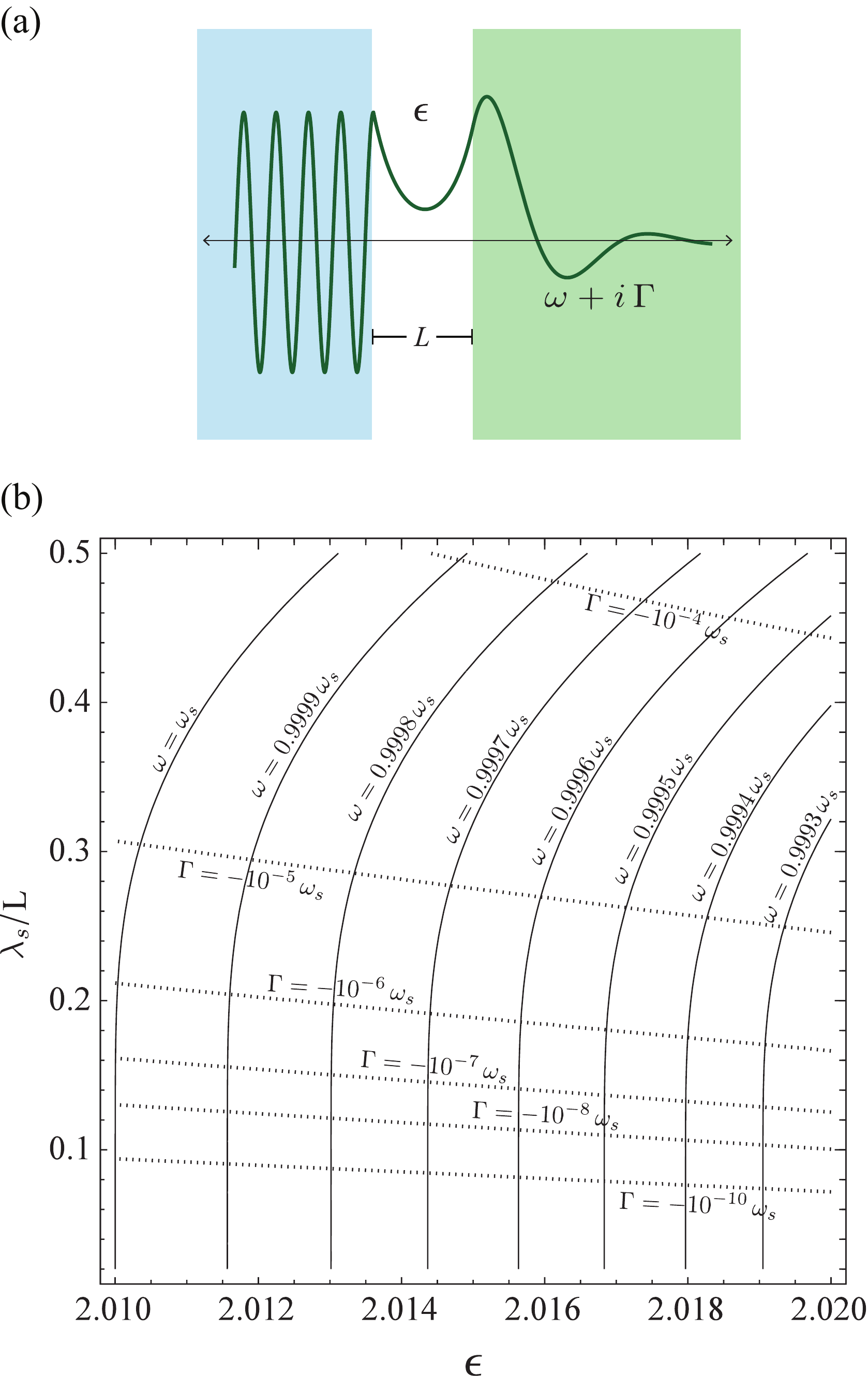}
\caption{\footnotesize The origin of exceptional point can be understood in terms of interaction of two halves of the sensor. One such half-system is shown by the schematic in (a) which includes an isotropic dielectric sandwiched between two semi-infinite media of a biaxial dielectric and a prism. The complex frequency $\omega + i \Gamma$ of the quasi-stationary ghost bound state depends on the permittivity $\epsilon$ and thickness $L$ of the isotropic layer as shown by the contour plots in (b).}
\label{Fig:single}
\end{figure}

In order to better understand the origin of exceptional point through the interaction of ghost waves and how the initial operating condition is achieved, let us consider ``half" of the sensor system which consists of an isotropic layer of finite thickness with a semi-infinite biaxial slab on one side and a prism on the other (see Fig. \ref{Fig:single}(a)). The frequency for the quasi-guided ghost surface mode for a given fixed tangential momentum will depend on both the permittivity $\epsilon$ and thickness $L$ of the isotropic layer as shown by the contour plots in Fig. \ref{Fig:single}(b). While the real part $\omega$ of the frequency indicates the resonance position in the reflection spectrum for the corresponding scattering system, the imaginary part $\Gamma$ is related to the resonance width and decay constant arising from the radiative leakage through the prism.\\

\begin{figure}[tb]
\centering
   \includegraphics[width=3.1 in]{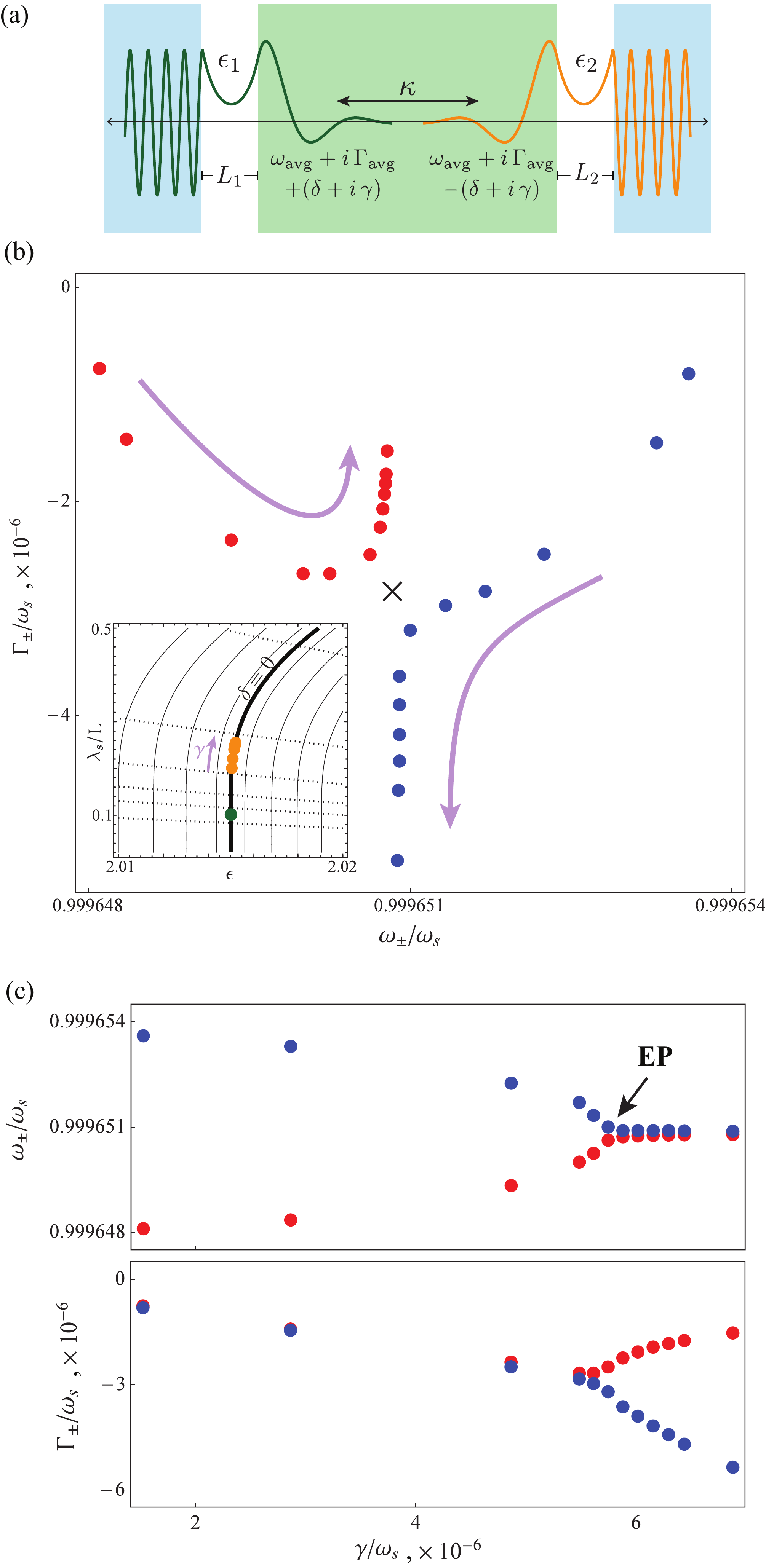}
\caption{\footnotesize Exceptional point arises from the interaction of ghost waves. (a) The schematic shows the two uncoupled ghost surface modes with a half-frequency difference of $\delta + i \gamma$ interacting with each other with a coupling strength $\kappa$. The resulting eigen-frequencies are shown in the complex plane in (b) as one changes the parameters $(\epsilon_2,L_2) $ of one of the half-systems which is indicated in the inset. Panel (c) shows the real and imaginary parts of these frequencies as a function of the decay contrast $\gamma$ which clearly indicates a phase transition through an exceptional point.}
\label{Fig:ep}
\end{figure}

The sensor geometry (see Fig. \ref{Fig:ep}(a)) can then be considered as a composite of two such half-systems (see Fig. \ref{Fig:single}(a)) each with its own parameters ($\epsilon_1, L_1$) and ($\epsilon_2, L_2$). If the interaction between the two modes, which is mediated by the finite thickness of the biaxial slab, is described by a coupling constant $\kappa$ (in frequency units) then the unperturbed composite system can be effectively described by a non-Hermitian Hamiltonian $\mathcal{H}$ of the form
\begin{equation}
\mathcal{H} = \omega_{\rm avg} +i\Gamma_{\rm avg} +
  \left( {\begin{array}{cc}
    \delta+i\gamma &  \kappa \\
   \kappa &  -\delta-i\gamma \\
  \end{array} } \right)
\end{equation}
where $\omega_{\rm avg} +i\Gamma_{\rm avg}$ is the average frequency for the two uncoupled quasi-bound states and $\delta +i\gamma$ is half the frequency difference. This effective Hamiltonian yields the system's eigen-frequencies as
\begin{equation}
\omega_{\pm} + i\Gamma_{\pm} = \omega_{\rm avg} +i\Gamma_{\rm avg} \pm \sqrt{(\delta+i\gamma)^2 + \kappa^2}.
\label{Eq:freqs}
\end{equation}

As one of the conditions to achieve exceptional point, we need to first have $\delta = 0$; that is, the two uncoupled modes need to have the same real parts for frequencies. Let us assume, without loss of generality, that ($\epsilon_1, L_1$) is fixed and we are free to vary ($\epsilon_2, L_2$). Then the condition corresponds to first identifying one of the contour lines for constant $\omega$, as those illustrated in Fig. \ref{Fig:single}(b), that goes through the parameters of the half-system described by ($\epsilon_1, L_1$). Any point on that contour line, as shown by the inset of Fig. \ref{Fig:ep}(b), can be assigned as the parameter value ($\epsilon_2, L_2$) of the other half-system so that the two uncoupled modes have the same real parts for frequencies ($\delta=0$) although the imaginary parts can be different ($\gamma \neq 0$).\\

The other condition for arriving at an exceptional point is to have $\gamma=\kappa$; that is, the imaginary parts of the uncoupled mode frequencies should have a half-difference $\gamma$ given by the coupling constant $\kappa$. This is achieved by moving along the contour line and picking up different values for ($\epsilon_2, L_2$) which will result in a continuous change in $\gamma$. Eventually, for one particular value of the parameters, the condition $\gamma=\kappa$ will be reached. The corresponding parameters will provide us with the exceptional point configuration for the composite system whose eigen-frequencies will be degenerate in both real and imaginary parts: $\omega_+=\omega_-$ and $\Gamma_+=\Gamma_-$, as can be readily seen from Eq. \ref{Eq:freqs} with these two conditions. \\

The process of finding the exceptional point is illustrated in Fig. \ref{Fig:ep}(b) which traces the location of the system's eigen-frequencies 
in the complex plane as the parameters of the half-system $(\epsilon_2, L_2)$ is varied along the contour as shown in the inset. Note that initially the two frequencies converge towards each other maintaining the same imaginary part. For one critical value of $(\epsilon_2,L_2)$, the frequencies reach other at an exceptional point which gives us the desired initial setting for sensing operation. If the parameters are varied further along the contour, the frequencies start to diverge away from each other, now maintaining a constant real part. \\

This converging and diverging behavior of eigen-frequencies is characteristic around an exceptional point which marks a phase transition that breaks parity-time ($\mathcal{PT}$) symmetry \cite{guo2009observation, ozdemir2019parity}. Since moving the parameter $(\epsilon_2,L_2)$ along the contour, with the direction shown in the inset of Fig. \ref{Fig:ep}(b), involves getting one of the prisms closer to the biaxial slab, it increases radiative leakage through that prism for one of the uncoupled modes, and hence increases their overall decay ``contrast" characterized by $\gamma$. Initially for small decay contrast $\gamma<\kappa$ when the mode interaction dominates, the eigenmodes for the composite system are effectively symmetric and anti-symmetric combinations of the two uncoupled modes which respect $\mathcal{PT}$ symmetry. Such hybridization of modes results in a splitting in the real parts for frequencies. However, since the two eignemodes are not localized to any one particular interface, they undergo the same decay, and therefore, the corresponding frequencies share the same imaginary part. \\

On the other hand, for large decay contrast $\gamma>\kappa$, the eigen-frequencies behave in the opposite manner. Here the modes couple too differently with the environment to couple strongly between themselves. Each eigenmode gets localized at one of the biaxial interfaces and breaks $\mathcal{PT}$ symmetry. Since the modes effectively do not interact with each other, their frequencies maintain the same real parts, while the corresponding imaginary parts are different as the modes undergo separate amounts of leakage on their respective sides.\\

The $\mathcal{PT}$ symmetric phase ($\gamma<\kappa$) and the broken $\mathcal{PT}$ phase ($\gamma>\kappa$), which are marked by the distinct behavior of their eigen-frequencies, are connected to each other at the exceptional point ($\gamma=\kappa$). At this transition point, the eigenmodes of the composite coalesce into one single state and the eigen-frequencies become degenerate in both real and imaginary parts. The whole phase transition process can be summarized in Fig. \ref{Fig:ep}(c) which is the signature for spontaneous breaking of $\mathcal{PT}$ symmetry at an exceptional point \cite{miri2019exceptional, guo2009observation}.\\

\begin{figure}[tb]
\centering
   \includegraphics[width=3.4 in]{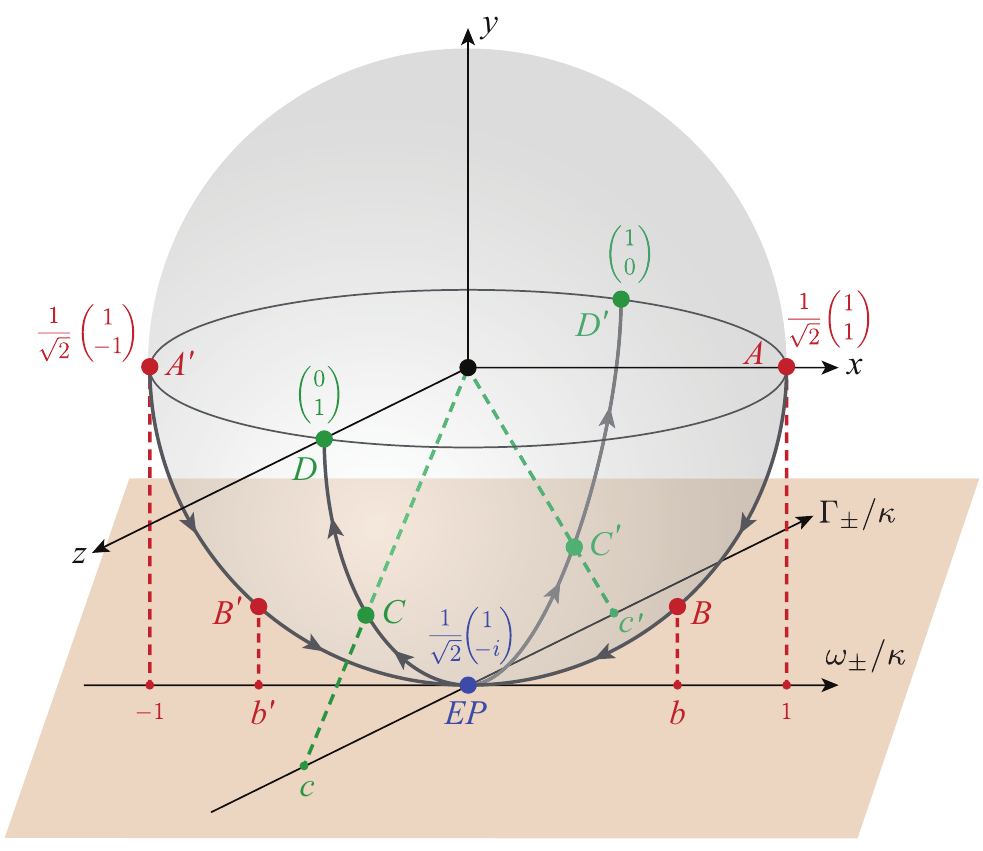}
\caption{\footnotesize A Bloch sphere can illustrate the behavior of eigenmodes and eigen-frequencies undergoing a $\mathcal{PT}$ symmetry breaking phase transition near a second-order exceptional point. Here the eigenmodes are represented by points on the Bloch sphere and the corresponding eigen-frequencies are projected onto the complex plane underneath it. The evolution starts from points $A$ and $A'$ which represent symmetric and antisymmetric combinations of the uncoupled states. As the decay contrast is increased, the eigenmodes move down towards the south pole while remaining on the $xy$ plane (points $B$ and $B'$) and their vertical projections onto the complex plane give the corresponding eigen-frequencies (points $b$ and $b'$). As the modes simultaneously arrive at the south pole, exceptional point is reached -- the modes coalesce into a single state and the eigen-frequencies become degenerate. If the decay contrast is further increased, the eigenmodes now move up towards the equator, this time being confined in the $yz$ plane (points $C$ and $C'$) with the eigen-frequencies given by their gnomonic projections (points $c$ and $c'$). The evolution continues with the eigenmodes asymptotically approaching the uncoupled states (points $D$ and $D'$). This geometric representation captures the trajectories of eigenmodes and eigen-frequencies in one coherent picture which also aid understanding of the initial biasing of the sensor.}
\label{Fig:bloch}
\end{figure}

To complete the discussion of how eigenmodes and eigen-frequencies behave near an exceptional point, Fig. \ref{Fig:bloch} illustrates their evolution by  representing the eigenmodes as points on the surface of a Bloch sphere and the corresponding eigen-frequencies on the complex plane underneath it. Here, the point where the complex plane touches the Bloch sphere at its south pole corresponds to $\omega_{\rm avg} +i\Gamma_{\rm avg}$.\\

When the decay contrast $\gamma=0$, the two eigenmodes are symmetric and antisymmetric combinations of the uncoupled states, and are therefore represented by points $A$ and $A'$ on the $x$ axis in Fig. \ref{Fig:bloch}. As $\gamma$ increases, for $\gamma < \kappa$, the eigenmodes move down towards the south pole while remaining on the $xy$ plane with the same $y$ coordinate (points $B$ and $B'$). Note that they are no longer antipodal points on the Bloch sphere -- which is characteristic of a non-Hermitian system. Their vertical projections on the complex plane below are the corresponding eigen-frequencies (points $b$ and $b'$), which approach each other with an increasing $\gamma$. \\

As we reach $\gamma=\kappa$, the two eigenmodes simultaneously arrive at the south pole coalescing to a single state. At this exceptional point, the two eigen-frequencies also become degenerate. As $\gamma$ is further increased, the eigenmodes now move up towards the equator maintaining the same $y$ coordinate, but this time being confined in the $yz$ plane (points $C$ and $C'$). The eigen-frequencies are given by the gnomonic projections \cite{coxeter1961introduction} of these points onto the complex plane from the center of the Bloch sphere (points $c$ and $c'$). Finally, as $\gamma \rightarrow \infty$, the eigenmodes asymptotically approach the uncoupled states represented by points $D$ and $D'$ on the $z$ axis while the difference between the imaginary parts of the eigen-frequencies grow without bound, as predicted by the simple two state model (see Eq. \ref{Eq:freqs}). This ``Bloch sphere on a complex plane" representation which illustrates the trajectories of eigenmodes and eigen-frequencies in one coherent picture helps us understand the initial ``biasing" of the proposed sensor.  \\

Once the sensor is set up in the exceptional point configuration, changes in permittivity of one of the isotropic sides induced by analyte binding will introduce a perturbation to the system Hamiltonian and the eigen-frequencies will no longer remain degenerate. As already well-established both theoretically \cite{kato2013perturbation, wiersig2014enhancing} and experimentally \cite{chen2017exceptional,hodaei2017enhanced}, the degeneracy splitting depends strongly on the amount of perturbation. The proposed ghost sensor shows such strong response as well (see Fig. \ref{Fig:sensor}(c)) which accounts for its high sensitivity.\\

Detecting the presence of analytes through the measurement of frequency shifts is a widely used method in optical biosensing \cite{mejia2018plasmonic}. As the proposed sensor operates by measuring the frequency splitting instead of single frequency shift, it has the additional advantage that the measurement process is intrinsically self-referenced \cite{zhu2010chip}; that is, there is no need for an external reference to account for thermal drifts, CCD variability etc.\\

\begin{figure}[tb]
\centering
   \includegraphics[width=3.2 in]{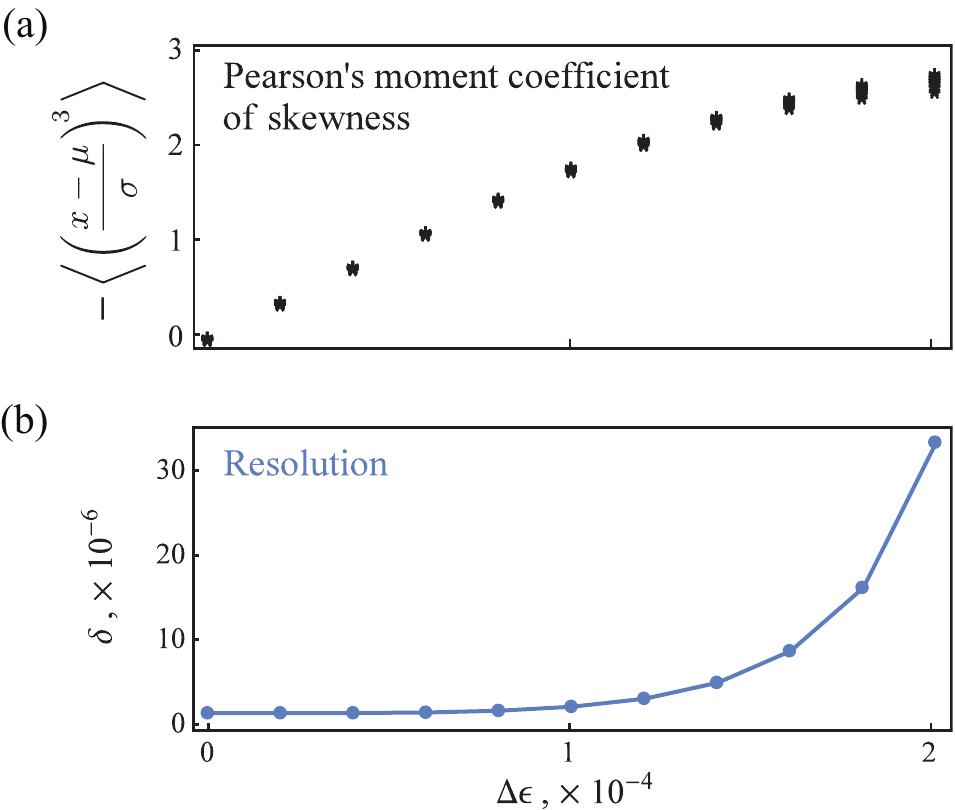}
\caption{\footnotesize Ghost sensor offers superior performance. Instead of detecting the individual frequency peaks, one can use a statistical measure of asymmetry of the detected spectrum, such as Pearson's moment coefficient of skewness, to determine the change in permittivity. The corresponding sensing performance shown in (a) is robust against noise. Here we used noise with a strength of 20 dB SNR and correlation frequency of $10^{-7} \omega_s$. Panel (b) shows the resolution $\delta$ of the sensor which reaches as fine as $10^{-6}$ RIU.}
\label{Fig:actual}
\end{figure}

A simpler sensor setup which would not require determination of exact frequency splitting can be realized by looking for the degree of asymmetry in the output spectrum of the sensor. A measure of Skewness by Pearson's moment coefficient \cite{joanes1998comparing}  (see Fig. \ref{Fig:actual}(a)) shows that by taking this third order moment of the spectrum one can determine analyte concentration quantitatively.\\

In practice, performance of any sensor is affected by noise, which can be of diverse origins, ranging from, for example, thermal fluctuations and detector dark current to high coherence of the illumination. As Fig. \ref{Fig:actual}(a) shows, the proposed sensor performs robustly in the presence of noise of varying strength and correlation time. As the sensor works by taking an average moment of the output spectrum, the effect of noise is substantially reduced.\\

Another metric for sensor performance is precision, or also known as sensor resolution which ultimately gives the limit of detection (LoD) of a sensor \cite{armbruster2008limit}. Since ghost sensor does not have any metallic components and deals with sharp resonances, the precision is as high as dielectric sensors based on waveguide interferometry.  Fig. \ref{Fig:actual}(b) shows the sensor precision as a function of the perturbation amount. Note that the precision is finest when the perturbation is minimum which promotes the sensor LoD.\\

To illustrate the superiority of ghost sensor over a plasmonic sensor of same level of design complexity, we investigate a surface plasmon resonance (SPR) sensor in Kretchmann configuration. 
Our calculation shows that ghost sensor offers orders of magnitude higher precision.\\

Ghost sensor can be realized with a variety of material selections.  In Ref. \cite{examples} we give a few examples of biaxial anisotropic materials that can be used as the core element of our proposed sensor. All these materials are low cost dielectric and are readily available, bringing the realization of the sensor well within experimental reach. Note that although biaxial materials have previously been used for sensing applications \cite{liu2011high}, our sensor is novel in that it is based on exceptional point operation and does not require any fabrication of chiral structures. \\  

Ghost sensing can also be implemented with one prism-coupler. If the second prism required to collect the transmission signal in the setup of Fig. \ref{Fig:sensor}(a) is removed, then the sensor can still work with the reflection spectrum. In this case the absorption of isotropic layer on the transmission side will play the role of leakage to set up the initial exceptional point configuration. Also, the sensor can work reasonably well even if it is not biased at an exceptional point. The sensitivity in that case is comparable with SPR sensors \cite{khan2021ghost}.\\

To conclude, we have developed a new approach to optical sensing with dramatically enhanced sensitivity. The sensor responsivity is improved due to the operation at exceptional point  that arises from the interaction between ghost surface waves. The proposed method does not require complex device geometry and precise gain-loss balance, thus offering the first practical approach to exceptional point-based sensors. The sensor employs low cost dielectric materials, offers high sensitivity and precision, shows robustness against noise and can accommodate compact miniaturization for point-of-care applications. Combining the extreme sensitivity with straightforward implementation in real-world environment, ghost sensors has the potential to make biosensing more accurate and more affordable.\\

The authors thank Prof. Mordechai Segev for helpful discussion and acknowledge support for this work from the the National Science Foundation, Grant No.  DMREF- 1629276, and the Gordon and Betty Moore Foundation.


%

\end{document}